\documentclass[conference]{IEEEtran}
\IEEEoverridecommandlockouts
\usepackage{cite}
\usepackage{amsmath,amssymb,amsfonts}
\usepackage{algorithmic}
\usepackage{graphicx}
\usepackage{textcomp}
\usepackage{xcolor}
\usepackage[colorinlistoftodos]{todonotes}
\usepackage{hyperref}
\usepackage[most]{tcolorbox}
\usepackage{subcaption}
\usepackage{xargs}
\usepackage{makecell}
\usepackage[algo2e]{algorithm2e}
\usepackage{algorithm}
\usepackage{algorithmic}
\usepackage{multirow}
\usepackage{multicol}
\usepackage{float}
\usepackage{orcidlink}

\newcommandx{\toDiscuss}[2][1=]{\todo[linecolor=red,backgroundcolor=red!25,bordercolor=red,#1]{#2}}
\newcommandx{\done}[2][1=]{\todo[linecolor=blue,backgroundcolor=blue!25,bordercolor=blue,#1]{#2}}

\definecolor{classColor}{HTML}{795E26}
\definecolor{typeColor}{HTML}{267F99}
\definecolor{keywordColor}{HTML}{0000FF}
\definecolor{mulColor}{HTML}{CC0000}

\newcommand{\twoShot}[0]{\texttt{2-shot}}
\newcommand{\threeShot}[0]{\texttt{3-shot}}

\hypersetup{pdfborder={0 0 0}}

\def\BibTeX{{\rm B\kern-.05em{\sc i\kern-.025em b}\kern-.08em
    T\kern-.1667em\lower.7ex\hbox{E}\kern-.125emX}}

\begin{document}

\title{Structure- and Event-Driven Frameworks for State Machine Modeling with Large Language Models}  

\author{
\IEEEauthorblockN{Samer Abdulkarim\textsuperscript{*}\IEEEauthorrefmark{2},
Evan Boyd\textsuperscript{*}\IEEEauthorrefmark{2},
Karl Bridi\textsuperscript{*}\IEEEauthorrefmark{2},
Alec Tufenkjian\textsuperscript{*}\IEEEauthorrefmark{2},
Boqi Chen\orcidlink{0000-0002-1451-3603}\IEEEauthorrefmark{2},
Gunter Mussbacher\orcidlink{0009-0006-8070-9184}\IEEEauthorrefmark{2}}
\IEEEauthorblockA{\IEEEauthorrefmark{2}\textit{Electrical and Computer Engineering, McGill University, Montreal, Canada} \\
\textit{Email: \{samer.abdulkarim, evan.boyd, karl.bridi, alec.tufenkjian, boqi.chen\}@mail.mcgill.ca, gunter.mussbacher@mcgill.ca}}

\thanks{\textsuperscript{*} All four authors contributed equally to this research.}

}

\maketitle

\begin{abstract}

UML state machine design is a critical process in software engineering for modeling dynamic system behavior. Traditionally, state machines are manually crafted by experienced engineers based on natural language (NL) requirements—a time-consuming and error-prone procedure. Many automated approaches exist but they require structured NL requirements. In this paper, we investigate the capabilities of current Large Language Models (LLMs) to \emph{fully automate} UML state machine generation via specialized \textit{State Machine Frameworks} (SMFs) from \textit{non-structured} NL requirements. We evaluate two types of state-of-the-art LLMs using single-step and multi-step prompting approaches: a non-reasoning LLM \texttt{GPT-4o} and a reasoning-focused LLM \texttt{Claude~3.5~Sonnet}, and introduce a novel \textit{Hybrid Approach} that uses the output from a \textit{Single-Prompt Baseline} as an initial draft state machine, which is then refined through an SMF. In our study, two distinct SMFs are developed based on human approaches: (i) a \textit{Structure-Driven SMF}, in which state machine components (states, transitions, guards, actions, etc.) are generated in sequential steps, and (ii) an \textit{Event-Driven SMF}, where identified events iteratively guide state machine construction. Our experiments indicate that while LLMs demonstrate a promising ability to generate state machine models from the \textit{Single-Prompt Baseline} (e.g., $F_1$-scores of 0.90 for states and 0.75 for transitions using \texttt{Claude~3.5~Sonnet}), their performance is not yet fully sufficient for a fully automated solution (e.g., $F_1$-scores of 0.23 for guards and 0.00 for actions for \texttt{GPT-4o}). Our proposed \textit{Hybrid Approach} improves the performance of the non-reasoning LLM (\texttt{GPT-4o}) to a similar level as the reasoning LLM (\texttt{Claude~3.5~Sonnet}) but does not further improve the reasoning LLM. Our evaluation highlights both the potential and the limitations of current LLMs for automated state machine design, providing a baseline for future research in this domain.
\end{abstract}


\section{Introduction}
\label{introduction}

\textit{Context:}
UML state machine design is an important process in software engineering that models the dynamic behavior of systems through states, transitions, and hierarchical features such as parallel regions, hierarchical states, and history states~\cite{uml2017}. Traditionally, state machines are manually crafted in a labor-intensive and error-prone process by experienced engineers who interpret natural language (NL) system descriptions to produce comprehensive models. Existing automated approaches require structured NL descriptions~\cite{cook1998discovering,ammons2002mining,lorenzoli2008automatic,biase2024}, motivating the need for novel approaches that can accurately capture system behavior from non-structured textual descriptions.

\textit{Problem Statement:}
In this paper, we address the problem of \emph{fully automated} state machine generation. Given non-structured NL system descriptions, our goal is to automatically derive a state machine that accurately represents the system’s behavior, including states, events, transitions, guards, actions, parallel regions, hierarchical states, and history states, without any human interaction.

Advancements in Large Language Models (LLMs) have shown they are capable of automating the modeling of many different types of software models, such as domain models~\cite{chen,camara2023assessment}, goal models~\cite{chen2023use}, and sequence diagrams~\cite{
jahan2024automated}. Although these LLMs demonstrate excellent understanding of input system descriptions, they still struggle to produce high-quality models fully autonomously~\cite{chen}. Multi-step prompting techniques such as Chain-of-Thought~\cite{wei2022chain} and Tree-of-Thought~\cite{yao2023tree} involve breaking down a complex problem into multiple sub-problems and invoking the LLM multiple times for each subtask. These strategies have been shown to improve LLM performance for automating modeling tasks~\cite{yang}.

Recently, emerging \emph{reasoning-focused LLMs} such as \texttt{DeepSeek-R1} 
and Anthropic’s \texttt{Claude 3.5 Sonnet} 
are designed to produce step-by-step logical reasoning in their outputs, demonstrating strong performance across challenging benchmarks. However, their effectiveness in software modeling tasks, especially for state machines, has not yet been systematically studied.

\textit{Objectives:}
Our study aims to evaluate the extent to which LLMs can be harnessed to fully automate the state machine design process using non-structured NL system descriptions. In particular, we investigate the single-step generation performance of two types of LLMs, a non-reasoning LLM and a reasoning LLM. Furthermore, we investigate (i) the effectiveness of different \emph{State Machine Frameworks} (SMFs), i.e., multi-step generation strategies, for generating state machines and (ii) the impact of a \textit{Hybrid Approach} combining single-step generation results and iterative refinement through SMFs.

\textit{Contributions:}
Specifically, this paper makes the following contributions to the modeling community:
\begin{itemize}
    \item We propose a fully automated pipeline for state machine generation that leverages LLMs through multi-step prompting. We present two distinct SMFs: a \textit{Structure-Driven SMF}, which generates state machine components in broad, sequential steps, and an \textit{Event-Driven SMF}, which iteratively guides generation based on identified events. Both SMFs are inspired by how humans approach the problem. We also explore a \emph{Hybrid Approach} that first generates an initial state machine using a \emph{Single-Prompt Baseline} and iteratively refines it through an SMF.

    \item We explore the impact of different reasoning strategies on state machine generation, comparing \texttt{GPT-4o} (a non-reasoning LLM) and \texttt{Claude 3.5 Sonnet} (a reasoning LLM) and discussing how their performance in state machine modeling varies.
        
    \item We present a comparative evaluation based on quantitative metrics (precision, recall, and $F_1$-score) obtained from a dataset of non-structured NL system descriptions with corresponding ground-truth state machines.
    
    \item Our experiment results provide insights into the strengths and limitations of current LLM-based approaches for state machine design, establishing a valuable baseline for future research in automated model generation.
\end{itemize}

\textit{Added Value:}
To the best of our knowledge, our work is the first LLM-driven effort to fully automate state machine generation from non-structured requirement descriptions. In this context, we aim to explore the use of multi-step generation strategies along with the differences between \emph{reasoning} and \emph{non-reasoning} LLMs, including a \textit{Hybrid Approach} that combines a \textit{Single-Prompt Baseline} with a multi-step generation strategy. 
A detailed comparison between the two types of LLMs opens new opportunities for integrating LLMs into model-driven engineering practices.
\section{Background}
\label{sec:background}

This section gives an overview of state machine modeling as well as different types of LLMs and prompting techniques.

\textbf{State Machine Modeling.} 
UML state machines~\cite{uml2017} are used to model reactive systems, capturing how a system transitions between states and performs actions in response to events and considering guard conditions. UML state machines extend basic state machines with features like hierarchically nested states (superstates), orthogonal regions for parallelism, and history states to remember prior substates, making them particularly suitable for specifying reactive and event-driven behaviors of software and embedded systems.
The significant expertise required for state machine modeling, which is also time-consuming and prone to human error, motivates automated state machine modeling: if non-structured NL requirements or scenarios could be translated into state machines automatically, it would save manual effort and reduce modeling errors.

\textbf{Reasoning vs. Non-Reasoning LLMs.}
LLMs differ in their ability to perform explicit reasoning. Some LLMs (e.g., \texttt{DeepSeek-R1}, 
\texttt{Claude 3.5 Sonnet}) 
are designed or fine-tuned to generate intermediate reasoning steps (a chain-of-thoughts) as part of their output. We refer to these as \textit{reasoning LLMs}. Other models (e.g., \texttt{GPT-4o}) 
typically provide direct answers without an explicit reasoning trace; we consider these \textit{non-reasoning LLMs}. \textit{Reasoning LLMs} have shown advantages in tasks requiring inference or complex problem solving~\cite{
ji-etal-2023-towards}. In the context of model generation, a reasoning LLM might explicitly perform a step-by-step analysis of the requirements, potentially yielding more coherent and complete models. While non-reasoning LLMs could generate explicit reasoning steps, they typically rely on external prompting strategies such as Chain-of-Thought~\cite{wei2022chain} prompting. In this work, we examine both reasoning and non-reasoning LLMs, exploring the possibility that LLMs designed for reasoning tasks, such as \texttt{Claude 3.5 Sonnet}, may require fewer carefully crafted prompts to achieve high-quality state machines.

\textbf{Prompting Techniques.}
Fine-tuning an LLM involves updating some of its parameters to fit a specific task and dataset. This practice is more common with smaller-scale LLMs. However, with larger LLMs ($>$ 1B Parameters), this approach becomes harder and less common due to training costs and lack of sufficient training data. Therefore, we adopt two alternative prompting techniques to help guide the LLM through the state machine generation process. \textbf{Few-shot Prompting}~\cite{brown2020} incorporates $N$ labeled examples into the task context alongside the task description and input. However, solely using input/output pairs in these examples may not suffice for the LLM to learn complex tasks that require multiple steps to solve. \textbf{Chain-of-Thought Prompting}~\cite{wei2022chain} addresses this issue by including a sequence of reasoning steps in the examples provided to the LLM. This approach enables the LLM to generate reasoning alongside the task output and has been shown to improve performance on various benchmarks compared to few-shot techniques.
We use a combination of both techniques to enhance performance.

\section{Approach}
\label{sec:approach}

In this section, we first describe our LLM framework for state machine modeling, which systematically converts a non-structured NL description of a reactive system into a UML state machine, and then discuss the investigated LLMs.

\subsection{Overview}
\begin{figure}[h]
\centering
\includegraphics[width=0.45\textwidth]{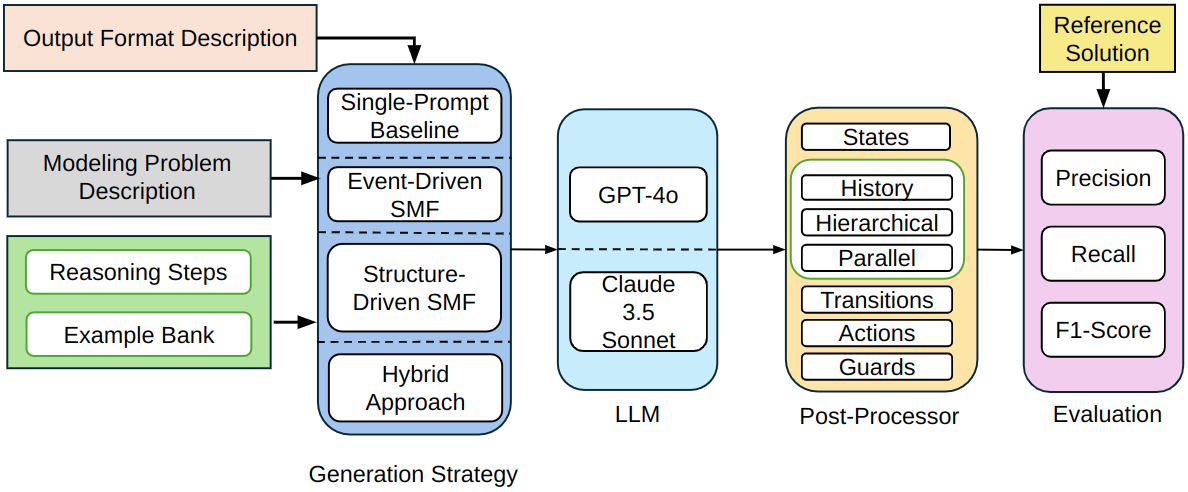}
\caption{Architecture for LLM-based state machine generation}
\label{fig:architecture}
\end{figure}

Figure~\ref{fig:architecture} provides an overview of the architecture. We focus on state machines with actions on transitions. The framework takes a \textit{modeling problem description} in natural language as input and produces a UML state machine model as output. At its core, an LLM (either \texttt{GPT-4o} or \texttt{Claude 3.5 Sonnet}) is controlled by one of four \emph{generation strategies} to guide the model through an incremental construction process. The generation process, specified and guided by the generation strategies, is divided into multiple stages, each corresponding to a certain aspect of the state machine, and is carried out in a step-by-step manner that mirrors the thought process of an experienced modeler (except for the \textit{Single-Prompt Baseline} generation strategy which is carried out in a single stage).

The \textit{Single-Prompt Baseline} prompts the LLM to directly produce an Umple~\cite{garzon2015umple} state machine, while the \textit{Structure-Driven SMF}, \textit{Event-Driven SMF}, and \textit{Hybrid Approach} prompt the LLM to generate the state machine in HTML table format to structure the LLM outputs across multiple steps. This process is formalized by considering a system description $d$ with an underlying ground truth state machine model $sm$ and defining a generator $\texttt{f}$ such that $sm' = \texttt{f}(d)$, where $sm'$ should ideally be similar to $sm$ (i.e., $sm' \cong sm$). Additionally, the framework includes a strict rule-based \emph{post-processor} module that merges and refines the LLM outputs. In case post-processing fails for a state machine component in a step, then the output of the last step where the component was successfully post-processed is considered. Finally, the \textit{evaluation} module measures the accuracy of generated state machines based on the ground truth (i.e., \textit{reference solution}).

\textbf{Single-Prompt Baseline} is a generation strategy with a 2-shot technique in which the full task is provided to the LLM in a single step. The prompt includes a description of the system and asks the LLM to output all states, transitions with guards and actions, hierarchical states, parallel regions, and history states while defining how those components play a role in building the state machine. Finally, we ask the LLM to produce a complete Umple state machine. This strategy leverages the LLM's ability to produce a full model in a single step.

\begin{figure}[h]
    \centering
    \includegraphics[width=0.45\textwidth]{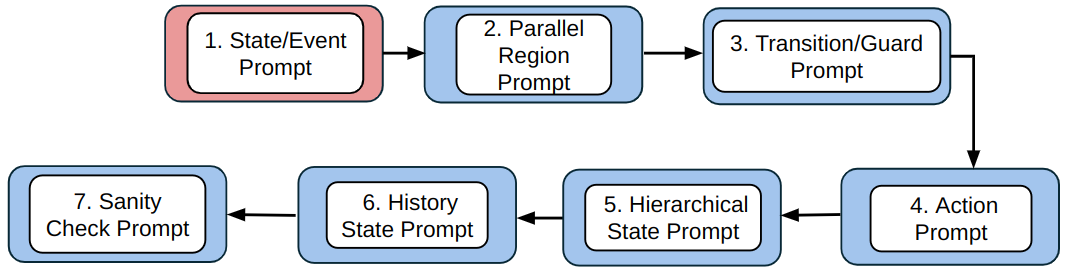}
    \caption{Structure-Driven State Machine Framework}
    \label{fig:linear-flow}
    \end{figure}

\begin{figure}[h]
    \centering
    \includegraphics[width=0.35\textwidth]{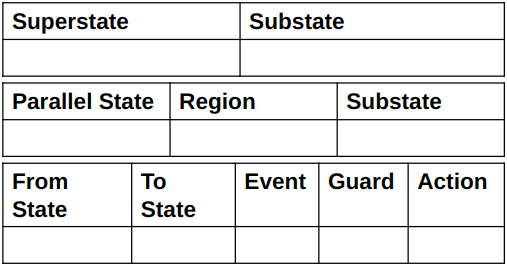}
    \caption{HTML table output for multi-step generation strategies}
    \label{fig:linear-representation}
    \end{figure}

\textbf{Structure-Driven SMF} is a multi-step, linear strategy that breaks down model generation by components in a linear sequence. The LLM is guided through a series of prompt stages, each focusing on a particular aspect of the state machine. The steps of this strategy are inspired by one way humans approach this task. \autoref{fig:linear-flow} illustrates the generation strategy's stages, each designed to find the core components of one aspect (for example, Step 1: ``List all relevant states and events for the system"; Step 2: ``Find all relevant states that undergo discrete events at the same time and factor them out into parallel region"; Step 3: ``For each identified state, list possible outgoing transitions. Specify guard conditions for transitions if applicable"; Step 4: ``Specify actions that occur on each transition"; and so on). After each stage, the partial output (as HTML tables) is fed into the next prompt so the LLM can build on it. This structure-driven generation strategy aims to allow the LLM to increase the accuracy of its generated state machine by tackling one state machine component at a time. Finally, the post-processor then outputs the state machine in three structured HTML tables that define all components of the state machine as shown in \autoref{fig:linear-representation}. 

     
\begin{figure}[t]
    \centering
    \includegraphics[width=0.45\textwidth]{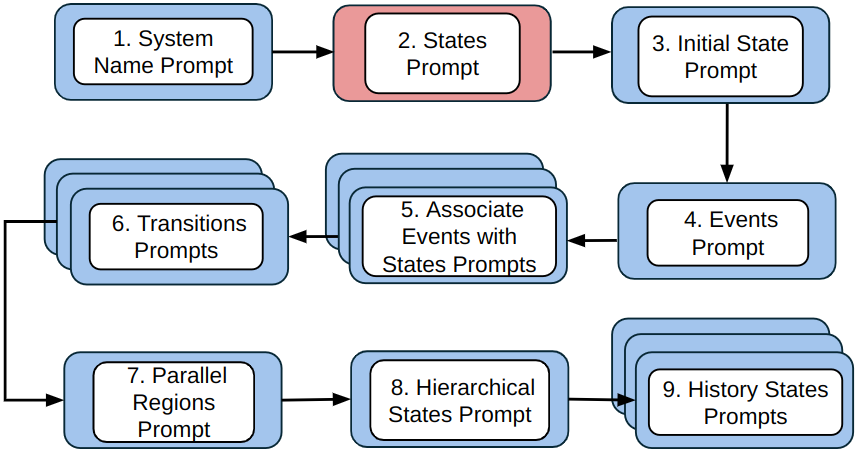}
    \caption{Event-Driven State Machine Framework}
    \label{fig:event-drivenframwork}
    \end{figure}

\textbf{Event-Driven SMF} is an iterative, multi-step generation strategy that structures the prompts around events, which are often used to distinguish states and define transitions in a state machine. Similar to the \textit{Structure-Driven SMF}, this strategy is inspired by an alternative way of how humans approach this task. \autoref{fig:event-drivenframwork} provides an overview of this approach. The approach first identifies the states to define the backbone or skeleton of the state machine and then extracts the relevant events from the input description. Then, for each event, it queries the LLM to determine how that event is handled in the state machine (for instance, Step 5: ``Consider the event X: describe the states where event X can occur"). The LLM responds with the partial state machine (states entered, exited, transitions taken, actions performed under that event). The framework accumulates these responses for all events, integrating them into a unified state machine. By focusing the LLM on one event at a time, this approach aims to improve recall of event-specific behaviors and ensure that no event is left unhandled. However, it requires merging partial models and may introduce duplicate or conflicting transitions that the post-processor must resolve. This approach guides the parallel state identification and the identification of hierarchical states as states that undergo common transitions. These common transitions are then grouped by the LLM to avoid duplication. The approach subsequently identifies history states by prompting the LLM to check events and states that may benefit from tracking history. Finally, the post-processor outputs the state machine in structured HTML tables (shown in \autoref{fig:linear-representation}), same as the \textit{Structure-Driven SMF}. 

\begin{figure}[tb]
    \centering
    \includegraphics[width=0.45\textwidth]{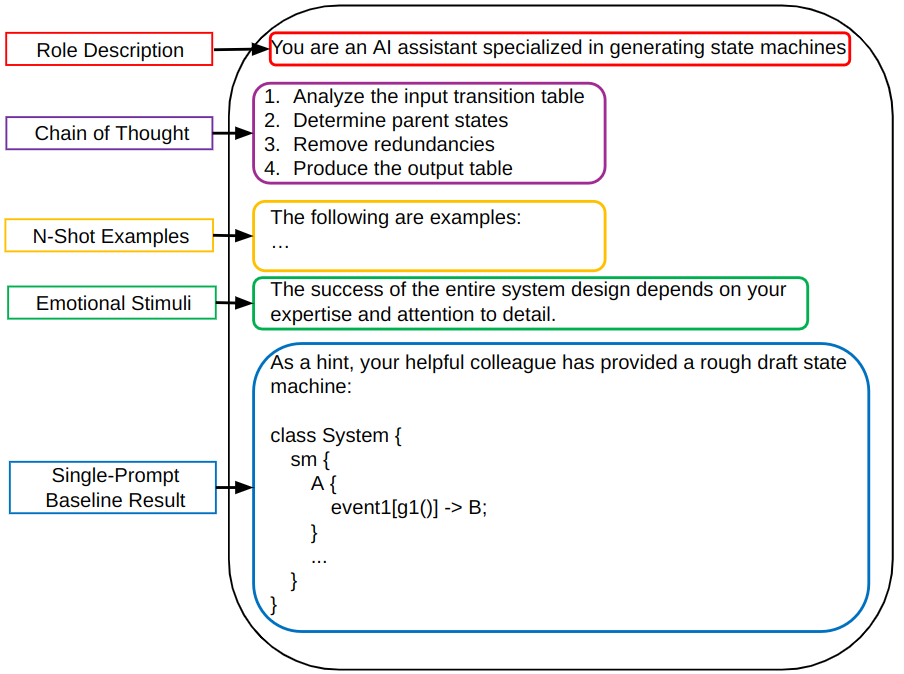}
    \caption{Prompt structure of the Hybrid Approach}
    \label{fig:combined-app}
    \end{figure}

\textbf{Hybrid Approach.} Although the SMFs decompose the state machine generation problem into multiple steps, each step focuses on a single aspect of the process, potentially leading to a loss of the overall context.
Thus, the \textit{Hybrid Approach} aims to take advantage of both the complete result from the \textit{Single-Prompt Baseline} and the step-by-step processes of the SMF approaches by merging the \textit{Single-Prompt Baseline} with the \textit{Structure-Driven SMF}.
Concretely, in the \textit{Hybrid Approach}, the LLM is first prompted to generate a complete state machine in Umple syntax using the \textit{Single-Prompt Baseline}. Then, this fully generated solution is appended at the end of all prompts within the \textit{Structure-Driven SMF} (e.g., ``this solution was provided by your helpful colleague as a baseline") to guide the LLM in refining and expanding upon the initial draft through the SMF. \autoref{fig:combined-app} illustrates the general prompt structure of the \textit{Hybrid Approach}, where the fully generated Umple syntax is appended to the end of a prompt used in the SMF.



\subsection{Large Language Models}
\label{sec:llm}
In our framework, we use two LLMs for state machine generation: the non-reasoning \texttt{GPT-4o}, 
developed by OpenAI, and the reasoning \texttt{Claude 3.5 Sonnet}, 
created by Anthropic. However, the framework is not limited and can be extended to other types of LLMs. 

\texttt{GPT-4o} has been shown to exhibit strong performance in tasks requiring mathematical reasoning, deductive logic, and algorithmic problem-solving, which may influence the conciseness and precision of its generated state definitions. Meanwhile, \texttt{Claude 3.5 Sonnet} has demonstrated competitive performance in tasks that require integrating broad knowledge with reasoning processes~\cite{huang2024olympicarenamedalranksintelligent}, such as cause-and-effect analysis and decompositional reasoning, which could explain \texttt{Claude 3.5 Sonnet}’s tendency to generate more detailed and context-rich descriptions. Our framework can take advantage of both by comparing outputs to a ground truth to help determine which LLM is better suited for automatically generating state machines. 

The temperature parameter~\cite{renze2024effect} is an important design decision for LLM-based state machine generation, as certain aspects of the generation involve more subjective judgment and can benefit from the creative capabilities of LLMs. 
Specifically, a higher-temperature setting can be useful for the
state generation steps. As many other steps rely on the generated states in the \textit{Structure-Driven SMF} and \textit{Event-Driven SMF}, it is important to maximize the recall of states by using the LLM's creative capabilities. Hence, we exploit the greater creativity that comes with higher temperatures for the steps highlighted in red in Figures~\ref{fig:linear-flow} and \ref{fig:event-drivenframwork}.


\begin{table}[b]
    \centering
    \caption{Summary of state machine components in ground-truth solutions (P..Printer, SM..Spa Manager, D..Dishwasher, C..Chess Clock, B..Bread Maker, T..Thermomix TM6, W..W-UMPLE, S..SSC7)}
    \begin{tabular}{|c|c|c|c|c|c|c|c|c|}
        \hline
         \textbf{\# Components} & P & SM & D & C & B & T & W & S \\
         \hline
        \textbf{States} & 6 & 11 & 9 & 9 & 9 & 9 & 17 & 7 \\
        \hline
        \textbf{Transitions} & 17 & 17 & 17 & 16 & 17 & 17 & 41 & 24 \\
        \hline
        \textbf{Guards} & 6 & 4 & 4 & 4 & 4 & 7 & 5 & 10 \\
        \hline
        \textbf{Actions} & 3 & 0 & 7 & 6 & 5 & 6 & 24 & 16 \\
        \hline
        \textbf{Hierarchical states} & 2 & 3 & 2 & 3 & 3 & 1 & 5 & 1 \\
        \hline
        \textbf{Parallel regions} & 0 & 5 & 2 & 2 & 0 & 0 & 2 & 0 \\
        \hline
        \textbf{History states} & 1 & 1 & 1 & 1 & 1 & 1 & 1 & 1 \\
        \hline
    \end{tabular}
    \label{tab:state_machine_example}
\end{table}


\section{State Machine Evaluation}
\label{sec:evaluation}

We discuss the evaluation procedure, scheme, and criteria.

\subsection{Evaluation Procedure}

To evaluate the state machine generation strategies proposed in this paper, we adopt an ideal scenario where the system description is clear and concise. This evaluation is based on eight modeling problem descriptions in English (see \autoref{tab:state_machine_example}) from an undergraduate university course, designed to assess students' proficiency in state machine design. Each problem comes with a reference state machine in diagrammatic format, created by modeling experts, which serves as the ground-truth for comparison. Our evaluation involves comparing the generated models to these reference solutions to assess the effectiveness of our strategies.

Due to the subjective nature of model-driven software design, two or more state machine designs may correctly model the same behavior for a given problem description. For instance, while state names, hierarchical state names, or parallel region names might differ between two state machines, the underlying transitions and overall behavior of the components could remain the same. Consequently, automating the evaluation of a state machine output is complex. Additionally, to the best of our knowledge, this is the first exploration into automated state machine generation from non-structured NL system descriptions with LLMs, and as such, no existing evaluators are available for performing automated state machine evaluation. Therefore, we manually evaluate the outputs from the generation strategies.




To streamline the evaluation process, we adopt an approach focused on efficiency and consistency. A single author conducts the evaluation for a given designed approach. The output syntaxes (Umple code for the \textit{Single-Prompt Baseline}, HTML tables for the other approaches) are compared against diagrammatic ground-truth solutions. To minimize bias, the evaluation protocol focuses on identifying exact or near-exact matches between the generated outputs and the ground-truth state machine. In essence, if two components are intended to represent the same concept (e.g., the same state or action), they are graded as equivalent, even if their names differ. This approach aims to provide a consistent and fair evaluation process across the generation strategies.

\subsection{Evaluation Scheme}
\label{subsec:evaluation_scheme}
When evaluating a state machine generated by the generation strategies, we considered the seven components of the state machine that we deemed to be the most characteristic or representative of the state machine modeling decisions. These seven components are the states
, transitions
, guards
, actions (
however, we are only considering actions on transitions and not entry/exit/do actions in states), hierarchical states
, parallel regions
, and history states
. The distribution of these components across the ground-truth state machines is shown in~\autoref{tab:state_machine_example}. Additionally, we also consider the aggregate of all components for one of the evaluation criteria as explained in ~\autoref{sec:quantitative}. For each component, we categorize the generated component depending on whether it has a match in the corresponding ground-truth model. The first category includes the generated components that have an exact match or semantic match with the ground-truth model (true positives). This also includes the components which are named differently than in the ground-truth model but serve the same purpose (superstates or parallel regions that contain the same set of matching substates for instance). The second category concerns generated elements that do not have an equivalent in the corresponding ground-truth model (false positives). By default, transitions that are connected to states that do not match any state in the ground-truth model are considered false positives. A similar strategy is applied to guards and actions that belong to false positive transitions. This underlines a strict evaluation schema for transitions, guards, and actions that recognizes the fact that they are of little value if the component(s) on which they depend are incorrect. 
The third and last category includes the components of the ground-truth model which do not have equivalents from the generated model (false negatives).

\subsection{Evaluation Criteria}
\label{sec:quantitative}

To evaluate the quality of the generated model, we compute the precision ($P = \frac{TP}{TP + FP}$), recall ($R = \frac{TP}{TP + FN}$), and $F_1$-scores ($F_1 = \frac{2 \times P \times R}{P + R}$) for each of the seven state machine components detailed in \autoref{subsec:evaluation_scheme} for each generated state machine, in addition to the overall result (computed by aggregating the true positives (TP), false positives (FP), and false negatives (FN) of all components). The overall $F_1$-score is calculated to estimate the overall quality of the output of a generation strategy. The individual metrics for the seven state machine components provide more detailed and interpretable results, as well as hints as to the next aspects to optimize with subsequent generation strategies. 
Precision gives an estimate of the accuracy of predictions made by the generation strategy.
Recall measures the ability of the generation strategies to find all relevant instances of the concerned component in the ground-truth model.
The $F_1$-score combines the precision and recall measures, i.e., representing the harmonic mean of precision and recall.


\section{Experiments}
\label{sec:experiment}

The aim of this section is to assess the capability of LLMs in state machine modeling to identify areas of the state machine modeling task where these models may encounter challenges, provide insights on the most effective state machine generation approach using such LLMs, and compare several LLMs. More concretely, we aim to investigate the following three research questions (RQ):

\begin{itemize}
\item[1:] How well do reasoning and non-reasoning LLMs generate state machines using a single-prompt technique?



\item[2:] How much do the multi-step generation strategies improve state machine generation for non-reasoning LLMs compared to the single-prompt technique from RQ1?



\item[3:] How well do the multi-step generation strategies from RQ2 generalize to reasoning LLMs for state machine generation?

\end{itemize}

We first describe the experimental setup for testing each approach in \autoref{sec:exp_setup}, followed by the experimental settings in \autoref{sec:exp_settings}. Subsequently, we tackle each one of the three research questions and report findings in Sections~\ref{sec:results-RQ1}, \ref{sec:results-RQ2}, and~\ref{sec:results-RQ3}, respectively. Finally, we provide an in-depth analysis and discussion of the results, including qualitative results in \autoref{sec:discussion} and state threats to validity in \autoref{sec:threats}.

\subsection{Dataset and Setup}
\label{sec:exp_setup}

To test each approach, we assembled a dataset of reactive system scenarios from projects and assignments in an undergraduate modeling course. Each scenario consists of a non-structured NL description of the system's behavior along with a manually crafted UML state machine diagram serving as the ground truth. The scenarios spanned various domains, such as a \textit{Dishwasher}, a \textit{Chess Clock}, a \textit{{Printer}}, and other common examples in which state machines are traditionally used. We ensured that some of the scenarios were sufficiently complex to identify the limitations of our generation approaches. In the end, the dataset contained eight unique system descriptions with varying complexity levels (see~\autoref{tab:state_machine_example}). 

\subsection{Experimental Settings}
\label{sec:exp_settings}

For each problem description, we ran the state machine generation using each of our generation strategies. We evaluated two LLMs: \texttt{GPT-4o} (accessed via OpenAI's API) and \texttt{Claude 3.5 Sonnet} (via Anthropic API). 
We experimented with two temperature settings:
\begin{itemize}
    \item \textbf{Deterministic setting:} $\text{temperature}=0.01$ to minimize randomness, so the LLM is likely to produce the most probable response, which we expected would reduce hallucinated model elements and improve precision.
    \item \textbf{Creative setting:} $\text{temperature}=0.5$, which injects moderate randomness, potentially increasing recall by allowing the model to explore less obvious possibilities (at the risk of adding incorrect elements).
\end{itemize}

We decided to use the deterministic setting for all steps in our generation strategies, except for ``State/Event Prompt" in \textit{Structure-Driven SMF} (\autoref{fig:linear-flow}), and ``States Prompt" in \textit{Event-Driven SMF} (\autoref{fig:event-drivenframwork}). We decided to use the creative setting for these steps as they benefit from exploring more possibilities. We also limited the LLMs to generate up to 1500 tokens for each prompt. 


\noindent
\noindent
\textbf{Prompt settings:} We use a \twoShot{} prompting strategy by selecting two examples from a pool of three state machines, including the \texttt{Printer}, \texttt{Spa Manager}, and \texttt{Dishwasher} examples, for our multi-step generation strategies. For our single-prompt strategy we employ \threeShot{} prompting, adding another ground truth state machine, \texttt{ChessClock}, to our pool, aiming to improve upon the baseline accuracy. Although these state machines can vary (e.g., some have more states or parallel regions), they are all of comparable complexity. If a particular state machine is used as test input, it is excluded from the examples shown to the LLM, ensuring that the LLM is never tested on a state machine it has already seen in the prompt. Otherwise, the first two or three state machines, respectively, are chosen for the prompt. 
For more details, the paper artifacts are available online~\cite{artifact_models2025}.

\subsection{RQ1: Performance of LLMs in State Machine Modeling}
\label{sec:results-RQ1}

RQ1 investigates the ability of LLMs to generate state machine models from a single prompt, focusing on the extent to which they can identify diverse state machine components as we are interested in how far the LLMs are from a perfect performance score for this task. 
Furthermore, we carry out a more fine-grained analysis by highlighting the modeling aspects the LLMs may struggle with. Particularly, we evaluate precision, recall, and $F_1$-scores when it comes to identifying states (including hierarchical and history states), transitions, guards, actions, and parallel regions.

\noindent
\textbf{Results.}
From the precision, recall, and $F_1$-scores for each state machine component in~\autoref{tab:single} for \texttt{Claude 3.5 Sonnet} and \texttt{GPT-4o}, we observe the following.

The best overall $F_1$-score is $0.7029$, achieved by \texttt{Claude 3.5 Sonnet}. \texttt{GPT-4o} obtains an overall $F_1$-score of $0.5431$, indicating that there is still room for improvement with single-step generation strategies.

Both LLMs perform comparatively well in detecting states (average $F_1$-scores of $0.8991$ with \texttt{Claude 3.5 Sonnet} and $0.8038$ with \texttt{GPT-4o}), suggesting that basic structure identification is relatively easier.

Performance for transitions is moderate (best $F_1$-score of $0.7502$ from \texttt{Claude 3.5 Sonnet}), but guards show a steeper performance drop (lowest $F_1$-score $=0.2348$ for \texttt{GPT-4o}), reflecting difficulties in extracting conditional logic.
Actions is the most challenging component to identify, with \texttt{Claude 3.5 Sonnet} yielding an $F_1$-score of $0.1633$ and \texttt{GPT-4o} failing to identify any actions ($F_1$-score $=0.000$).

On one hand, \texttt{Claude 3.5 Sonnet} achieves a moderate $F_1$-score for hierarchical states (0.6486) and parallel regions (0.5333), but a poor value for history states (0.2500). In contrast, \texttt{GPT-4o} performs moderately in hierarchical states (0.5810) and history states (0.4286), but poorly in parallel regions (0.1905). Overall, both LLMs struggle with capturing advanced state machine components.

\begin{table}[t]
    \centering
    \caption{Performance metrics for Single-Prompt Baseline}
    \begin{tabular}{|c|c|c|c|}
        \multicolumn{4}{c}\textbf{Claude 3.5 Sonnet} \\
        \hline
        \textbf{Component} & \textbf{Precision} & \textbf{Recall} & \textbf{$F_1$-score} \\
        \hline
        States & \textbf{0.8994} & \textbf{0.9053} & \textbf{0.8991} \\
        \hline
        Transitions & \textbf{0.8057} & \textbf{0.7062} & \textbf{0.7502} \\
        \hline
        Guards & \textbf{0.7244} & \textbf{0.5206} & \textbf{0.5645} \\
        \hline
        Actions & \textbf{0.2857} & \textbf{0.1143} & \textbf{0.1633} \\
        \hline
        Hierarchical states & 0.5646 & \textbf{0.8333} & \textbf{0.6509} \\
        \hline
        Parallel Regions & \textbf{0.4500} & \textbf{0.7000} & \textbf{0.5333} \\
        \hline
        History States & 0.2500 & 0.2500 & 0.2500 \\
        \hline
        All & \textbf{0.7931} & \textbf{0.6384} & \textbf{0.7029} \\
        \hline
        \multicolumn{4}{c}\textbf{GPT-4o} \\
        \hline
        \textbf{Component} & \textbf{Precision} & \textbf{Recall} & \textbf{$F_1$-score} \\
        \hline
        States & 0.8737 & 0.7552 & 0.8038 \\
        \hline
        Transitions & 0.6702 & 0.5206 & 0.5741 \\
        \hline
        Guards & 0.3095 & 0.2044 & 0.2348 \\
        \hline
        Actions & 0.0000 & 0.0000 & 0.0000 \\
        \hline
        Hierarchical states & \textbf{0.7143} & 0.5000 & 0.5810 \\
        \hline
        Parallel Regions & 0.3333 & 0.1333 & 0.1905 \\
        \hline
        History States & \textbf{0.4286} & \textbf{0.4286} & \textbf{0.4286} \\
        \hline
        All & 0.7129 & 0.4501 & 0.5431 \\
        \hline
    \end{tabular}
    \label{tab:single}
\end{table}

From these observations, we note that \texttt{Claude 3.5 Sonnet} outperforms \texttt{GPT-4o} across almost all state machine components in a single-prompt scenario. Moreover, both LLMs exhibit higher precision than recall on average (see ``All’’ rows in Table~\ref{tab:single}), suggesting that while the LLMs correctly identify several state machine components, they often fail to capture all relevant instances. As a result, the overall $F_1$-scores are substantially penalized by the lower recall values.

\begin{tcolorbox}
\textbf{Answer to RQ1.} While LLMs (particularly the reasoning LLM, \texttt{Claude 3.5 Sonnet}) demonstrate a promising ability to generate state machines from a single prompt, their performance is not yet sufficient for fully automated generation. The greatest challenges lie in identifying actions and more complex structures such as parallel regions and history states. Precision is often better than recall, indicating that the missing elements negatively affect the overall $F_1$-score. Hence, considerable scope remains for improvement in generating complete and accurate state machine models.
\end{tcolorbox}

\subsection{RQ2: Multi-Step Approaches for Non-Reasoning LLMs}
\label{sec:results-RQ2}

The generated state machines are influenced by the corresponding generation strategy. RQ2 investigates, for the non-reasoning LLM \texttt{GPT-4o}, (1) whether more granular generation pipelines help produce better state machines and (2) if the \textit{Hybrid Approach} using the output of the \textit{Single-Prompt Baseline} is better than the \textit{Single-Prompt Baseline} on its own. 
Concretely, we compare the average overall $F_1$-scores on various state machine components for each generation strategy: the \textit{Single-Prompt Baseline}, \textit{Structure-Driven SMF}, \textit{Event-Driven SMF}, and \textit{Hybrid Approach}. For the \textit{Hybrid Approach}, we combine the \textit{Single-Prompt Baseline} and the \textit{Structure-Driven SMF}, because the \textit{Structure-Driven SMF} clearly outperforms the \textit{Event-Driven SMF} (see~\autoref{tab:gpt4rq2}).


\begin{table*}[t]
\caption{Comparison of $F_1$-scores for state machine components 
for \texttt{GPT-4o}}
\centering
\begin{tabular}{|c|c|c|c|c|c|c|c|}
\hline
Generation Strategy & States & Transitions & Guards & Actions & Hierarchical States & Parallel Regions & History States \\
\hline
Single-Prompt Baseline & 0.8038 & 0.5741 & 0.2348 & 0.0000 & 0.5810 & 0.1905 & \textbf{0.4286} \\
Structure-Driven SMF & 0.7377 & 0.6277 & 0.2611 & 0.3250 & 0.6962 & 0.1429 & 0.1250 \\
Event-Driven SMF & 0.6584 & 0.3432 & 0.2295 & 0.2391 & 0.6208 & 0.3173 & 0.2083 \\
Hybrid Approach & \textbf{0.8582} & \textbf{0.7107} & \textbf{0.4240} & \textbf{0.3436} & \textbf{0.7928} & \textbf{0.3429} & 0.1250 \\
\hline
\end{tabular}
\label{tab:gpt3rq2}
\end{table*}


\begin{table}[t]
\caption{Comparison of overall average precision, recall, and $F_1$-scores 
for \texttt{GPT-4o}}
\centering
\begin{tabular}{|c|c|c|c|}
\hline
Generation Strategy & Precision & Recall & $F_1$-Score \\
\hline
Single-Prompt Baseline & \textbf{0.7130} & 0.4501 & 0.5431 \\
Structure-Driven SMF & 0.6562 & 0.6268 & 0.6260 \\
Event-Driven SMF & 0.2667 & \textbf{0.6870} & 0.3735 \\
Hybrid Approach & 0.7110 & 0.6142 & \textbf{0.6559} \\
\hline
\end{tabular}
\label{tab:gpt4rq2}
\end{table}

\noindent
\textbf{Results.} From the $F_1$-scores for each state machine component in~\autoref{tab:gpt3rq2} and the overall precision, recall, and $F_1$-scores in~\autoref{tab:gpt4rq2}, we observe the following 
for \texttt{GPT-4o}.

The \textit{Hybrid Approach} (0.6559) achieves the best overall $F_1$-score, followed by the \textit{Structure-Driven SMF} (0.6260). Both outperform the \textit{Single-Prompt Baseline} (0.5431), while the \textit{Event-Driven SMF} is lowest overall (0.3735).

The \textit{Hybrid Approach} yields the highest $F_1$-score for states (0.8582), outperforming the \textit{Single-Prompt Baseline} (0.8038). Both the \textit{Structure-Driven SMF} (0.7377) and \textit{Event-Driven SMF} (0.6584) show lower performance in identifying states.

The \textit{Hybrid Approach} (0.7107) performs substantially better than the \textit{Single-Prompt Baseline} (0.5741) in identifying transitions. The \textit{Structure-Driven SMF} (0.6277) also show improvements over the baseline.
The \textit{Hybrid Approach} (0.4240) performs best for guards, then the \textit{Structure-Driven SMF} (0.2611), the \textit{Single-Prompt Baseline} (0.2348), and the \textit{Event-Driven SMF} (0.2295).
While the \textit{Single-Prompt Baseline} completely fails to identify actions (0.0000), all multi-step generation strategies show substantial improvements  (0.3436 for \textit{Hybrid Approach}, 0.3250 for \textit{Structure-Driven SMF}, 0.2391 for \textit{Event-Driven SMF}). 


\begin{table*}[t]
\caption{Comparison of $F_1$-scores for state machine components 
for \texttt{Claude 3.5 Sonnet}}
\centering
\begin{tabular}{|c|c|c|c|c|c|c|c|}
\hline
Generation Strategy & States & Transitions & Guards & Actions & Hierarchical States & Parallel Regions & History States \\
\hline
Single-Prompt Baseline & \textbf{0.8991} & \textbf{0.7502} & \textbf{0.5645} & 0.1633 & 0.6509 & 0.5333 & 0.2500 \\
Structure-Driven SMF & 0.8203 & 0.5145 & 0.2744 & 0.2380 & 0.5592 & 0.5333 & 0.1250 \\
Event-Driven SMF & 0.7314 & 0.2988 & 0.1525 & 0.1862 & 0.4750 & \textbf{0.5500} & \textbf{0.4583} \\
Hybrid Approach & 0.8737 & 0.7209 & 0.4152 & \textbf{0.3375} & \textbf{0.7132} & 0.3939 & 0.3333 \\
\hline
\end{tabular}
\label{tab:claude_component_rq3}
\end{table*}

\begin{table}[t]
\caption{Comparison of overall average precision, recall, and $F_1$-scores 
for \texttt{Claude 3.5 Sonnet}}
\centering
\begin{tabular}{|c|c|c|c|}
\hline
Generation Strategy & Precision & Recall & $F_1$-Score \\
\hline
Single-Prompt Baseline & \textbf{0.7931} & 0.6384 & \textbf{0.7029} \\
Structure-Driven SMF & 0.5041 & 0.5116 & 0.5026 \\
Event-Driven SMF & 0.2038 & \textbf{0.6542} & 0.3052 \\
Hybrid Approach & 0.6368 & 0.6473 & 0.6336 \\
\hline
\end{tabular}
\label{tab:claude_overall_rq3}
\end{table}

The \textit{Hybrid Approach} performs best (0.7928) for hierarchical states, followed by the \textit{Structure-Driven SMF} (0.6962). Both outperform the \textit{Event-Driven SMF} (0.6208). The \textit{Single-Prompt Baseline} shows the lowest performance (0.5810).

The \textit{Hybrid Approach} excels in identifying parallel regions (0.3429), followed closely by the \textit{Event-Driven SMF} (0.3173). Both generation strategies improve on the \textit{Single-Prompt Baseline} (0.1905) and the \textit{Structure-Driven SMF} 0.1429.
The \textit{Single-Prompt Baseline} performs best in identifying history states (0.4286), followed by the \textit{Event-Driven SMF} (0.2083), and the \textit{Hybrid Approach} and \textit{Structure-Driven SMF} achieving the same results (0.1250).

The \textit{Hybrid Approach} performs best overall and for six out of seven state machine components. The \textit{Single-Prompt Baseline} is out-performed for a state machine component by a multi-step generation strategy in 13 out of 21 cases; the notable exception being history states for which the \textit{Single-Prompt Baseline} performs best. 

In general, precision is higher than recall for all generation strategies, except for the \textit{Event-Driven SMF} which performs very poorly for precision, resulting in its low $F_1$-score. Compared to the \textit{Single-Prompt Baseline}, the precision results are lower for the \textit{Structure-Driven SMF} (-0.0568 from 0.7130 to 0.6562) and almost identical for the \textit{Hybrid Approach} (0.7130 vs. 0.7110), indicating the advantage of the initial complete solution. On the other hand, the recall results are substantially higher for the \textit{Structure-Driven SMF} and the \textit{Hybrid Approach} (+0.1767 from 0.4501 to 0.6268 and +0.1641 from 0.4501 to 0.6142, respectively). Therefore, the performance gains of these two multi-step generation strategies over the \textit{Single-Prompt Baseline} can be attributed to their ability to identify a larger number of state machine components.

\begin{tcolorbox}
	\textbf{Answer to RQ2.} The \textit{Structure-Driven SMF} and \textit{Hybrid Approach} have overall positive effects on generated state machine quality for the non-reasoning LLM \texttt{GPT-4o}, due to substantial improvements in recall (+0.1641 to +0.1767). Although the \textit{Hybrid Approach} achieves the best overall performance, close to the reasoning LLM \texttt{Claude 3.5 Sonnet}, this improvement is not consistent across all state machine components
    , suggesting that an optimal approach might involve further generation strategy combinations.
\end{tcolorbox}


\subsection{RQ3: Multi-Step Approaches for Reasoning LLMs}
\label{sec:results-RQ3}

RQ3 aims to investigate the generalizability of the multi-step generation strategies (\textit{Structure-Driven SMF}, \textit{Event-Driven SMF}, and \textit{Hybrid Approach}) from RQ2 for the reasoning LLM \texttt{Claude 3.5 Sonnet}. While the \textit{Structure-Driven SMF} and \textit{Hybrid Approach} improved performance for the non-reasoning LLM \texttt{GPT-4o}, it remains unclear whether the same trend holds for reasoning LLMs. Hence, we compare the performance of \texttt{Claude 3.5 Sonnet} when guided by the various generation strategies. Like RQ2, we examine the overall precision, recall, and $F_1$-score for generated state machines, and the component-wise $F_1$-score.

\noindent
\textbf{Results.} From the $F_1$-scores for each state machine component in Table~\ref{tab:claude_component_rq3} and the overall precision, recall, and $F_1$-scores in Table~\ref{tab:claude_overall_rq3}, we observe the following for our generation strategies for \texttt{Claude 3.5 Sonnet}.

The best overall precision and $F_1$-score is still achieved by the \textit{Single-Prompt Baseline}, while overall recall is improved by the \textit{Event-Driven SMF} and the \textit{Hybrid Approach} compared to the \textit{Single-Prompt Baseline}. The \textit{Single-Prompt Baseline} achieves an overall $F_1$-score of 0.7029, substantially outperforming the \textit{Hybrid Approach} (0.6336), \textit{Structure-Driven SMF} (0.5026), and \textit{Event-Driven SMF} (0.3052). This pattern is inconsistent with our findings for the non-reasoning LLM \texttt{GPT-4o}, where the \textit{Hybrid Approach} yields the best results (0.6559), followed by the \textit{Structure-Driven SMF} (0.6260), with the \textit{Single-Prompt Baseline} achieving only 0.5431. In the case of the reasoning LLM \texttt{Claude 3.5 Sonnet}, the multi-step generation strategies do not make up substantial precision losses with substantial recall gains.

For individual state machine components, the results are more varied for the reasoning LLM \texttt{Claude 3.5 Sonnet}. The \textit{Single-Prompt Baseline} delivers the highest $F_1$-scores for three out of seven state machine components: states (0.8991), transitions (0.7502), and guards (0.5645). However, the \textit{Hybrid Approach} performs best for actions (0.3375) and hierarchical states (0.7132) and the \textit{Event-Driven SMF} performs best for parallel regions (0.5500) and history states (0.4583).

Comparative studies on LLM code generation capabilities have shown that \texttt{Claude 3.5 Sonnet} and earlier Anthropic LLMs outperform GPT models in specialized coding tasks \cite{Sobo31122025,wang2024aigeneratedcodereallysafe,murr2023testingllmscodegeneration,nascimento2024llm4dsevaluatinglargelanguage}. This aligns with our findings, where \texttt{Claude 3.5 Sonnet} achieves the highest overall $F_1$-score of 0.7029 with the \textit{Single-Prompt Baseline}---a single-step strategy that generates Umple state machine code. 
By contrast, the multi-step strategies primarily guide an LLM through the reasoning process of state machine design, with HTML tables serving as structured output for organizing state machine components rather than a mechanism for direct code generation. The lower overall $F_1$-scores of the \textit{Hybrid Approach} (0.6336), \textit{Structure-Driven SMF} (0.5026), and \textit{Event-Driven SMF} (0.3052) suggest that the multi-step strategies do not leverage a key strength of \texttt{Claude 3.5 Sonnet}: its single-step code generation capabilities. Moreover, multi-step strategies may interfere with the step-by-step reasoning process embodied within reasoning LLMs such as \texttt{Claude 3.5 Sonnet}.

\begin{tcolorbox}
\textbf{Answer to RQ3.}  
For the \texttt{Claude 3.5 Sonnet} reasoning LLM, the \textit{Single-Prompt Baseline} achieves a higher overall state machine $F_1$-score than the \textit{Structure-Driven SMF}, \textit{Event-Driven SMF}, and \textit{Hybrid Approach}. Furthermore, the \textit{Single-Prompt Baseline} obtains the best $F_1$-score for three of the seven state machine components, suggesting that the multi-step generation strategies from RQ2 do not consistently translate to improved performance when applied to a reasoning LLM.
\end{tcolorbox}

\subsection{Discussion}
\label{sec:discussion}


\noindent
\textbf{RQ1.} 
While \texttt{Claude 3.5 Sonnet} scores better on most state machine components than \texttt{GPT-4o}, both LLMs struggle with specific components, particularly actions, parallel regions, and history states. The precision-recall gap observed in both LLMs (with precision typically larger than recall) indicates that the LLMs prefer producing a conservative set of elements they are confident about, rather than attempting to capture all possible ones. The poor performance on actions highlights a significant limitation in current LLMs' ability to extract non-explicit behaviors from textual descriptions. Similarly, the challenges with advanced structural state machine components like parallel regions and history states suggest the need for more advanced generation strategies that can better guide LLMs through the modeling process.

\noindent
\textbf{RQ2.}
Our experiments reveal that more granular generation strategies substantially improve state machine generation quality for the non-reasoning LLM \texttt{GPT-4o}. The \textit{Structure-Driven SMF} and \textit{Hybrid Approach} both show substantial improvements over the \textit{Single-Prompt Baseline}. This suggests that breaking down the complex task of state machine design into more manageable subtasks helps non-reasoning LLMs better understand and model the problem domain. Furthermore, the \textit{Hybrid Approach} captures the straightforward comprehensiveness of the \textit{Single-Prompt Baseline} while benefiting from the more methodical element-by-element breakdown of the granular strategies. The poor performance of the \textit{Event-Driven SMF} compared to other strategies is unexpected, given that state machines are inherently event-driven. One possible explanation is that while this strategy achieves the highest recall, it suffers from extremely low precision, indicating that it tends to overgenerate elements that are not present in the ground truth. Interestingly, different generation strategies show varied strengths across different components. For the multi-step generation strategies, the \textit{Hybrid Approach} performs best for all state machine components except for history states, for which the \textit{Event-Driven SMF} performs best.
The most notable improvement across strategies was seen in actions, which was substantially improved in the \textit{Hybrid Approach}. However, even with these improvements, the $F_1$-scores remain relatively low, indicating that this remains a challenging aspect of state machine modeling for LLMs.

\noindent
\textbf{RQ3.} Contrary to our initial expectations based on the results from RQ2, we found that the \textit{Single-Prompt Baseline} achieves the best performance with the reasoning LLM \texttt{Claude 3.5 Sonnet} compared to the more complex multi-step generation strategies we developed. The findings from RQ3 highlight a key distinction between non-reasoning and reasoning LLMs in how they handle the various generation strategies. The non-reasoning LLM \texttt{GPT-4o} benefits from following the \textit{Structure-Driven SMF} and \textit{Hybrid Approach} multi-step generation strategies. On the other hand, none of the multi-step generation strategies improve overall performance over the \textit{Single-Prompt Baseline} for \texttt{Claude 3.5 Sonnet}, i.e, a single LLM call with a carefully crafted prompt achieves the best overall performance of all experiments. However, two of the multi-step generation strategies improve performance for specific state machine components: the \textit{Hybrid Approach} performs best for actions ($F_1$-score $= 0.3375$) and hierarchical states ($F_1$-score $= 0.7132$), while the \textit{Event-Driven SMF} was most effective for modeling parallel regions ($F_1$-score $= 0.5500$) and history states ($F_1$-score $= 0.4583$). These findings suggest that while non-reasoning LLMs benefit from multi-step generation strategies, such strategies may interfere with the inherent step-by-step reasoning process of reasoning LLMs. Thus, exploring alternative approaches to further enhance the performance of reasoning LLMs for state machine generation is necessary.

\subsection{Threats to Validity}
\label{sec:threats}

\noindent
\textbf{Internal validity.}
Due to the temperature of 0.5 used in the creative setting for state generation steps (\autoref{sec:exp_settings}), the generated state machines can be different for each run of the \textit{Structure-Driven SMF}, \textit{Event-Driven SMF}, and \textit{Hybrid Approach}. Based on informal observations, we believe the higher temperature is of value as it leads to more states being proposed by the LLM, thus improving recall rates. We mitigate this risk by averaging the results of eight state machine examples. However, determining the optimal temperature is out of scope for this paper, and other temperature settings may lead to different results. The manual evaluation of the experiments is done by a subset of the authors, which may introduce bias. We compensate for this bias by agreeing on evaluation guidelines. The strictness of the evaluation regarding transitions, guards, and conditions may lead to lower evaluation results but ensures greater consistency across different graders.


The \textit{Single-Prompt Baseline} and the multi-step generation strategies use different output syntaxes to represent state machines. The \textit{Single-Prompt Baseline} leverages the LLM's ability to generate a state machine in a single step, producing Umple code in outputs. Conversely, the multi-step generation strategies generate HTML tables for structured intermediary representations and to facilitate smoother post-processing. This fundamental difference means that the \textit{Single-Prompt Baseline} focuses on direct code generation, whereas the multi-step generation strategies are structured reasoning tasks. As such, performance results should be considered in the context of these fundamental differences in task structure and output representation. Furthermore, the strict post-processor module for HTML tables may suppress valid LLM outputs that are not fully compliant, hence influencing the final result. 

\noindent
\textbf{External validity.}
We experiment on one reasoning-focused LLM (\texttt{Claude 3.5 Sonnet}) and one non-reasoning LLM (\texttt{GPT-4o}), which is a limited set of LLMs, so the results might not generalize to all types of reasoning and non-reasoning LLMs. We compensated for the limited set of LLMs by choosing LLMs in both categories that demonstrated competitive performance in the aspects of interest, as explained in \autoref{sec:llm}. To address the absence of benchmark datasets for LLM-based generation of state machines that include both non-structured NL problem descriptions and reference solutions, we have compiled a dataset consisting of eight state machine modeling examples, each accompanied by reference models developed by modeling experts. However, our dataset of eight examples is a limitation. While these examples vary in complexity, they come from an undergraduate modelling course and may not generalize when applied to different domains or more complex systems.

\noindent
\textbf{Construct validity.}
We use recall, precision, and $F_1$-score which are standard for evaluating generative models~\cite{chen}.
\section{Related Work}
\label{sec:rel-work}



\textbf{Automated state machine generation.} 
Early studies primarily rely on statistical methods to extract state machines from structured software logs and execution traces~\cite{cook1998discovering,ammons2002mining,lorenzoli2008automatic}. More recent approaches incorporate various machine learning techniques, such as active learning, to improve extraction quality and efficiency~\cite{isberner2015open,yang2019improving}. With the rise of LLMs, Wei et al.~\cite{wei2024inferring} propose ProtocolGPT which analyzes protocol source code to identify and reconstruct underlying state machines.
However, existing approaches predominantly focus on extracting state machines from source code or structured data (e.g., Given-When-Then~\cite{biase2024}). In this paper, we explore the potential of LLMs to generate state machines directly from non-structured NL descriptions.

\textbf{LLMs for Model Generation.}
A frequent application of LLMs in MDE is the automatic generation of models from NL text~\cite{di2025use}. Early studies typically use various prompting techniques, relying exclusively on a single LLM call to generate complete models, including domain models~\cite{chen,arulmohan2023extracting}, goal models~\cite{chen2023use}, sequence diagrams~\cite{
jahan2024automated}, and use case models~\cite{tabassum2024using}. 
More recently, many multi-step model-generation methods based on iterative LLM interactions have been proposed, e.g., to integrate human best-practice modeling processes with iterative LLM-based domain model generation~\cite{yang}. Additionally, the Chain-of-Layers approach~\cite{zeng2024chain} utilizes a top-down iterative approach, combining LLMs with masked language models to progressively construct taxonomies. This paper is an application of LLMs for model generation, focusing on state machines with single-step and multi-step generation strategies. 


\textbf{LLMs for MDE.} 
Besides complete model generation, LLMs have been integrated into various other aspects of MDE. A common use case is model completion, where LLMs assist human modelers by providing completions or suggestions during the modeling process~\cite{camara2023assessment,chaaben2023towards
}. This approach targets more efficient and accurate model construction. 
Aligned with recent trends toward multi-step generation approaches~\cite{yang}, this paper investigates two \textit{State Machine Frameworks} inspired by how humans approach state machine modeling. Additionally, we compare generation outcomes using non-reasoning and reasoning-based LLMs, analyzing their relative strengths and effectiveness.
\section{Conclusion}
\label{sec:conclusion}

We present an LLM-based framework for the automated generation of UML state machines from non-structured NL descriptions. We introduce and compare three multi-step generation strategies—the \textit{Structure-Driven State Machine Framework (SMF)}, \textit{Event-Driven SMF}, and \textit{Hybrid Approach}—against the \textit{Single-Prompt Baseline}. While both tested LLMs, \texttt{GPT-4o} and \texttt{Claude 3.5 Sonnet}, show a promising ability in generating state machines using the \textit{Single-Prompt Baseline} (with \texttt{Claude 3.5 Sonnet} achieving the best overall result of all experiments), their performance remains insufficient for a fully automated solution. For the non-reasoning LLM (\texttt{GPT-4o}), the \textit{Structure-Driven SMF} and \textit{Hybrid Approach} improve state machine quality due to minor losses in precision but substantial gains in recall. However, for the reasoning LLM (\texttt{Claude 3.5 Sonnet}), the multi-step generation strategies do not consistently outperform the \textit{Single-Prompt Baseline}, suggesting that improvements in performance from our multi-step generation strategies do not always generalize to reasoning LLMs.


This study demonstrates the potential of LLMs for fully automated state machine modeling while also identifying key challenges that must be overcome for full automation. Future work includes investigating further generation strategies to improve low-performing state machine components (i.e., predominantly actions, parallel regions, and history states), how to best combine the strengths of different generation strategies, developing generation strategies tailored for reasoning LLMs, evaluating additional reasoning and non-reasoning LLMs for state machine generation, and developing a larger benchmark dataset for UML state machine generation.



\bibliographystyle{IEEEtran}
\bibliography{main}

\end{document}